\documentclass[12pt]{article}
\pdfoutput=1
\pdfoutput=1

\usepackage{graphicx}
\usepackage{psfrag}
\usepackage{enumerate}
\usepackage{soul}
\usepackage{subfig}
\usepackage{jheppubcustom}

\newcommand{\be}{\begin{equation}}
\newcommand{\ee}{\end{equation}}
\newcommand{\bea}{\begin{eqnarray}}
\newcommand{\eea}{\end{eqnarray}}

\def\[{\left [}
\def\]{\right ]}
\def\({\left (}
\def\){\right )}

\def\r2{\sqrt{2}}

 \def\simleq{\; \raise0.3ex\hbox{$<$\kern-0.75em
      \raise-1.1ex\hbox{$\sim$}}\; }
   \def\simgeq{\; \raise0.3ex\hbox{$>$\kern-0.75em
      \raise-1.1ex\hbox{$\sim$}}\; }

\newcommand{\bbibitem}[1]{\bibitem{#1}\marginpar{#1}}

\newcommand{\figref}[1]{Fig. \ref{#1}}
\newcommand{\secref}[1]{Sec. \ref{#1}}

\def\Label#1{\label{#1}%
  \smash{\hbox to0pt{\raise1ex\hbox{\tiny[#1]}\hss}}}
\def\noLabels{\let\Label=\label}
\def\nobbibitem{\let\bbibitem=\bibitem}

\begin{document}

\noLabels
\nobbibitem

\DeclareGraphicsExtensions{.pdf,.png,.gif,.jpg,.eps}

\title{ Spatial Curvature Falsifies Eternal Inflation}

\author{Matthew Kleban and Marjorie Schillo}

\emailAdd{mk161@nyu.edu}
\emailAdd{mls604@nyu.edu}

\affiliation{\it Center for Cosmology and Particle Physics\\
New York University \\
New York, NY 10003, USA}

\abstract{Inflation creates  large-scale cosmological density perturbations that are characterized by an isotropic, homogeneous, and Gaussian random distribution about a locally flat background.  Even in a flat universe, the spatial curvature measured within one Hubble volume receives  contributions from long wavelength perturbations, and will not in general be zero.   These same perturbations determine the Cosmic Microwave Background (CMB) temperature fluctuations, which are ${\cal O}(10^{-5})$. Consequently, the low-$l$ multipole moments in the CMB temperature map  predict the value of the measured spatial curvature $\Omega_k$.   On this basis we argue that a measurement of $|\Omega_{k}| >  10^{-4}$  would rule out slow-roll eternal inflation in our past with high confidence, while a measurement of $\Omega_{k} < -10^{{-4}} $ (which is \emph{positive} curvature, a locally closed universe) rules out false-vacuum eternal inflation as well, at the  same confidence level. In other words, negative curvature (a locally open universe) is consistent with false-vacuum eternal inflation but not with slow-roll eternal inflation, and positive curvature falsifies both.   Near-future experiments will dramatically extend the sensitivity of $\Omega_k$ measurements and constitute a sharp test of these predictions.  
}

\maketitle
	
\begin{table}[!ht]
\begin{tabular}{|l|l|}
 \hline
 \multicolumn{2}{|c|}{\bf  Table of symbols and definitions} \\
  \hline
    SREI, FVEI & Slow roll eternal inflation, false vacuum eternal inflation\\
   $\phi, V(\phi)$ & Inflaton field, inflationary potential\\
 $H_i, H_0$ & Hubble parameter during inflation, today\\
  $\Phi_N$ & Scalar perturbation in Newtonian gauge\\
  $g({\bf k}), {\mathcal P}({\bf k})$ & Gaussian distributed random variable and its power spectrum\\
   $k_* (=2 \pi/L_{*})$ & Momentum (length) scale corresponding to eternal inflation ($\delta \rho/\rho \sim 1$)\\
  $\beta$ & ${\cal O}(1)$ amplitude of ${\mathcal P}({\bf k})$ for $k < k_*$ \\
   $\langle X \rangle_g, X_{{av}}$ & Average of $X$ over $g$, spatial average of $X$ over  last scattering volume \\
 $ {\mathcal R}^{(3)}, \Omega_k$ & Ricci scalar of spatial slice, fraction of energy density in curvature\\
  $a_{lm}, C_{l}$ & Multipole moments of  CMB temperature, moments averaged over $m$\\
  $T_{0}=2.73K$ & Average CMB temperature \\ 
      
  \hline \end{tabular}
\end{table}

\section{Introduction} \label{intro}

Inflation was introduced as a mechanism to explain the observed homogeneity and isotropy of the universe \cite{Guth:1980zm}.    As a bonus, it naturally provides a nearly scale-invariant spectrum of primordial density perturbations that fit observational data very well.  For these reasons it has been widely accepted as a part of the standard cosmological model.

While only approximately 60 e-foldings\footnote{The precise number depends logarithmically on several poorly constrained factors, including the reheating temperature.}  of inflation are necessary to explain the data, many models of inflation can support eternal inflation (\figref{pots}).  As explained in more detail in \secref{ei}, eternal inflation predicts that the vast majority of the volume in the universe will inflate forever at a very rapid rate, with localized pockets or bubbles occasionally exiting from this phase.  The pockets may continue to undergo (ordinary, non-eternal) inflation for some time, after which they reheat and continue with standard post-inflationary cosmological evolution.  Even if the classical initial conditions do not fall in the eternally inflating regime of the parameter space of a given model, quantum mechanically it is impossible to set the weight of such configurations to zero.  Because eternal inflation produces arbitrarily large amounts of volume, arguably we should expect arbitrarily large numbers of e-folds of inflation in our past in \emph{any} model that supports eternal inflation in \emph{any} region of its configuration space.\footnote{The validity of this point of view remains very uncertain due the so-called ``measure problem"; see  \cite{Freivogel:2011eg} for a recent discussion.}  This includes most effective field theories of slow-roll inflation, and almost certainly string theory.

In this note we demonstrate that eternal inflation is directly falsifiable by observations that will be made in the next few years.  We argue that an observation of \emph{positive} spatial curvature significantly above the level of the cosmic microwave background (CMB) temperature fluctuations falsifies with very high confidence eternal inflation in our immediate past.  Negative spatial curvature at the same level falsifies slow-roll eternal inflation in our past.   In symbols, and bearing in mind the standard convention that $\Omega_k < 0$ corresponds to \emph{positive} spatial curvature (closed universe): $\Omega_k \simleq -10^{-4}$ falsifies all models of eternal inflation, while  $\Omega_k \simgeq 10^{-4}$ falsifies slow-roll eternal inflation.  Because observations currently constrain $|\Omega_k| \simleq 10^{-2}$ \cite{Komatsu:2010fb}, and near-future experiments will dramatically improve this bound \cite{Mao:2008ug, Barenboim:2009ug}, this constitutes a sharp test  of eternal inflation models.  Indeed, at least with certain choices of measure nearly \emph{all} inflationary models predict eternal inflation in our past, making this a very powerful check of the inflationary paradigm.

Our argument is based on the fact that the measured value of $\Omega_k$ is determined by almost exactly the same metric fluctuations that determine the low-$l$ multipole moments of the CMB temperature map, especially the $l=2$ quadrupole moments.  This holds true in any model in which large-scale densities fluctuations can be treated as statistically homogeneous, isotropic, and Gaussian around a flat background.  In particular, modes with a wavelength significantly longer than the Hubble length affect both the quadrupole moments and the measured $\Omega_k$ equally.  Because the quadrupole moments of the CMB are $\sim 10^{-5}$, it is extremely unlikely in such a model that the spatial curvature is significantly larger.

The reader can be forgiven for wondering why the title of this paper is not a tautology.  After all, doesn't inflation inflate away curvature, and eternal inflation most of all?  Not quite:  inflation produces fluctuations in the curvature, and during eternal inflation these are  large.  Other readers may wonder the opposite, why it is permissible to expand around a  flat slicing.  In fact, in universes undergoing eternal inflation, the large density fluctuations on large scales ensure that no globally well-defined notion of spatial curvature exists.  Slices of constant density may not remain spacelike, or they may ``terminate'' at future infinity, thereby dividing the universe into isolated pockets.\footnote{Completely generally, since we can only see a finite portion of the universe, any discussion of the global spatial curvature always requires either an extrapolation or an assumption. }

Nevertheless, within any approximately homogeneous and isotropic volume, the average spatial curvature is a measurable quantity---but it should be thought of as a local quantity related to the density within that volume rather than as a global, intrinsic characteristic of the universe as a whole.  A long period of inflation in the past (either truly eternal or simply significantly longer than the 60 e-folds required to solve the flatness problem) guarantees that any memory of the initial conditions---including any initial spatial curvature---is not detectable, and what remains are fluctuations generated during inflation.\footnote{Even when the global curvature is well-defined and non-zero, it can still be treated as a linear perturbation around a flat background on length scales less than the radius of curvature.}  Observations constrain the curvature within our last scattering volume to be $|\Omega_k| \simleq 10^{-2}$, well within the linear regime.  Therefore, for the purposes of studying local quantities (like the observed curvature or CMB multipole moments), we can treat the perturbations generated during slow-roll eternal inflation as   Gaussian fluctuations around a flat background.  As we will discuss below, the case of false vacuum eternal inflation requires a slightly separate treatment---not because memory of the initial conditions remains, but because the exit from the eternal phase is non-Gaussian.

\paragraph{Relation to previous work:}

Several analyses have focused on the connection between large-scale fluctuation modes and the low-$l$ moments of the CMB, beginning with \cite{GZ}.  The authors of \cite{Knox:2005hx, Waterhouse:2008vb, Vardanyan:2009ft} studied the variance of $\Omega_k$ to determine the in-principle precision with which it can be measured, and commented on the tension between spatial curvature and inflation.  The issue of negative curvature in universes that tunneled from a false vacuum was considered in \cite{Garriga:1998px, Freivogel:2005vv}.

\section{Eternal inflation} \label{ei}

\begin{figure}
\begin{center}$
\begin{array}{| c| c |}
\hline
	\includegraphics[angle=0, width=0.48\textwidth]{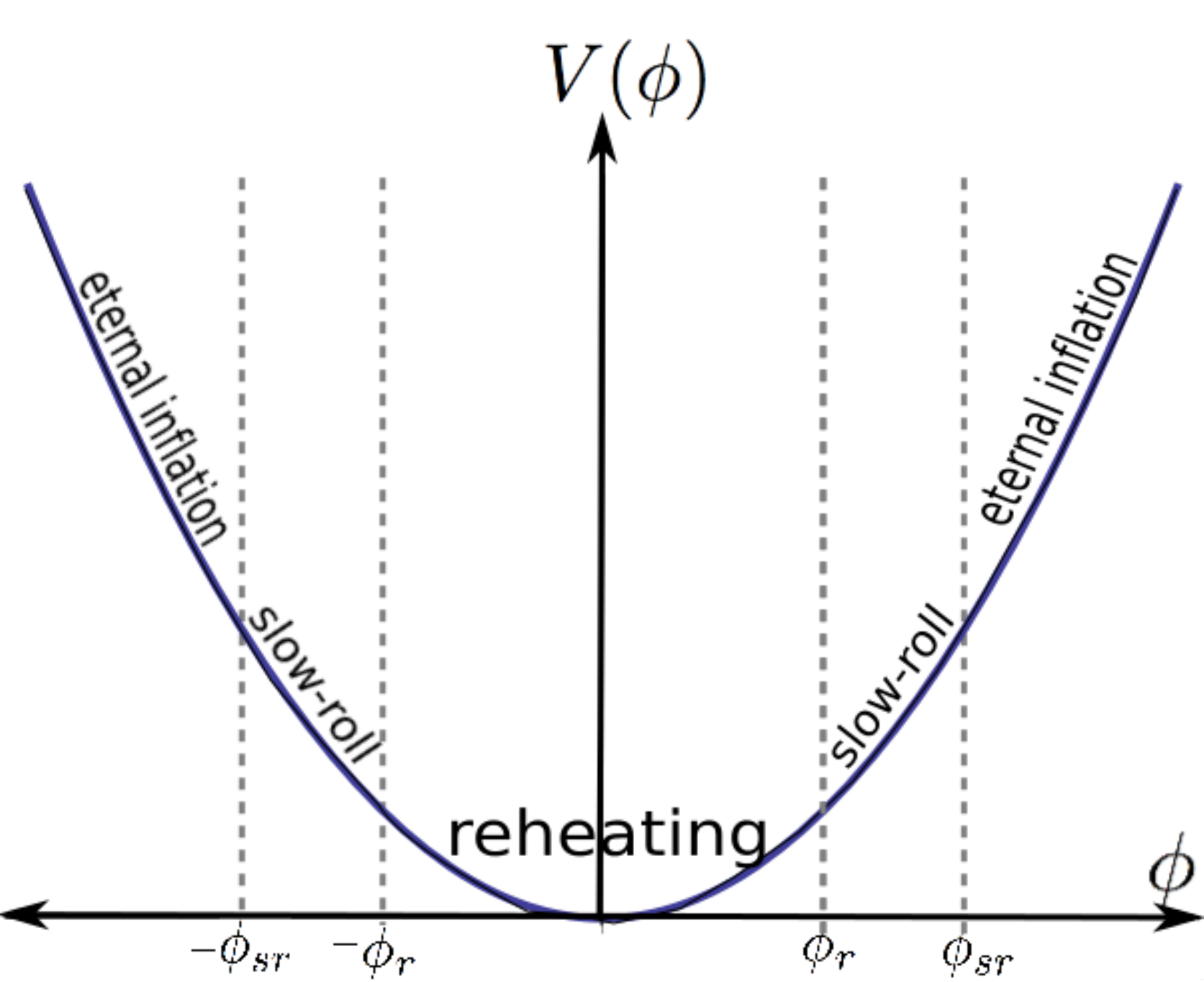} & \includegraphics[angle=0, width=0.5\textwidth]{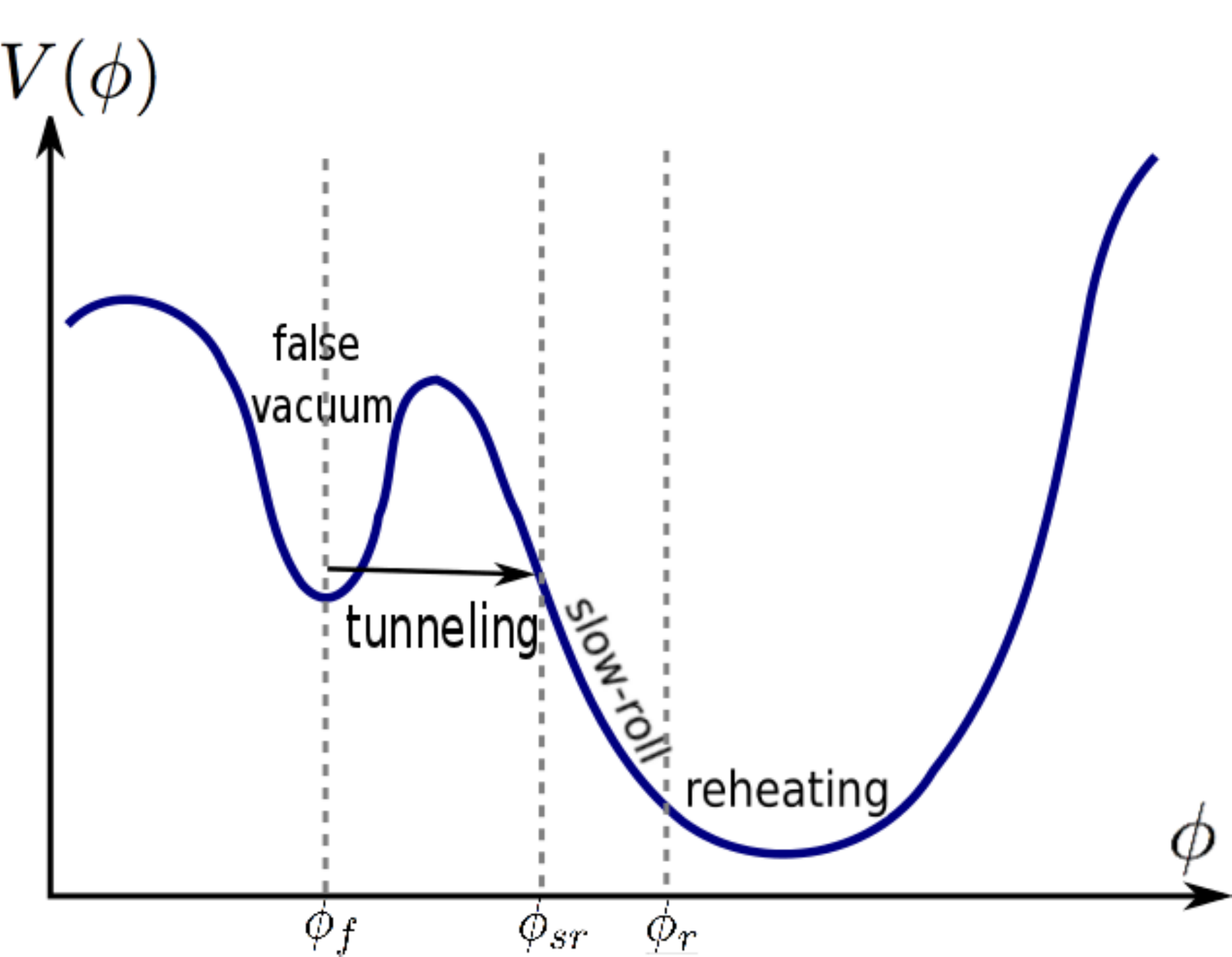} \\
		\hline
\end{array}$
\end{center}
\caption{\label{pots} Left panel: An inflationary potential $V(\phi) = \alpha \phi^n$ that supports both ordinary ($\delta \rho/\rho \ll 1$) and eternal ($\delta \rho/\rho \simgeq 1$) slow roll inflation.   Right panel: An inflationary potential suitable for false vacuum eternal inflation (FVEI).  Locally, the false vacuum decays by bubble formation, and the bubble undergoes a period of ordinary slow roll inflation after it forms.}
\end{figure}

Eternal inflation (see \cite{Guth:2007ng} for a recent review and further references) comes in two flavors:  ``slow roll eternal inflation'' (SREI) in which the inflaton field rolls  down a smooth, slightly tilted potential, and ``false vacuum eternal inflation'' (FVEI) where the inflation takes place in a false minimum with non-zero potential energy, and  ends locally through a first-order phase transition---the nucleation of a bubble.   In both forms of eternal inflation arbitrarily large amounts of volume are produced, and the typical region in which observers may find themselves may be expected to have a very large number of e-folds of inflation in its past.  We will take a large number of past e-folds of inflation as a defining property of eternal inflation; as a consequence, no memory of initial conditions can remain.

The operational definition of SREI is a period of inflation during which the typical quantum fluctuation in the inflaton,  $\delta \phi \sim H$, is of order the classical roll  $\Delta \phi \sim \dot \phi/H$.  Because $\delta \rho/\rho \sim H^2/{\dot \phi}$, this is equivalent to requiring that $V(\phi)$ and the background field value $\phi$ satisfy
\be \label{srei}
  {V(\phi)^{3/2} \over  \partial_\phi V(\phi)} M_P^{-3} \sim \delta \rho/\rho \simgeq 1.
  \ee

FVEI takes place in a minimum $V'(\phi_{f})=0$ such that $V(\phi_{f})>0$ (\figref{pots}).  Inflation takes place due to the constant positive value of $V$ in this minimum, and comes to an end locally via bubble nucleation.  Before taking perturbations into account, each bubble contains a homogeneous and isotropic Friedmann-Robertson-Walker (FRW) cosmology with negative curvature ($\Omega_k > 0$).  A \emph{flat} spatial slicing of such a bubble has large density perturbations on large scales because it cuts through the bubble, its walls, and extends into the FVEI region outside.  Density perturbations on spatial slices with \emph{negative} curvature do not have this characteristic---they can remain small on all scales, because a certain set of negatively curved slices fit nicely inside the bubble.  However, \emph{positive} curvature is even more difficult to achieve through random fluctuations here than it is in SREI; an observation of positive curvature would falsify FVEI at equal or greater confidence as compared to SREI.

In all types of inflation the value of the field is related to length scale---early times during inflation correspond to values of the inflaton far from the end of the inflationary plateau and hence to scales that exited the horizon early and are now very large, while late times correspond to values close to the end and hence to scales that exited the horizon late and are now smaller.  In all models of eternal inflation, the density perturbations on spatially \emph{flat} slices are ${\cal O}(1)$ on sufficiently large scales (corresponding to the eternal part of the inflaton potential, \eqref{srei}, or the false vacuum in FVEI).

Once the inflaton exits the eternally inflating phase, either through tunneling or  by classical rolling, a period of ordinary slow-roll inflation may begin.  Indeed, such a phase is required observationally.  For SREI, the absence of such a phase would mean that density perturbations are ${\cal O}(1)$ on all scales,  in conflict with observation.  Instead, Grishchuk  and Zel'dovich showed long ago \cite{GZ} that ${\cal O}(1)$ density perturbations can exist only on length scales much larger than the horizon size, since otherwise the CMB quadrupole would be too large.    For FVEI,  the bubble is dominated by negative curvature when it first forms.  Absent a period of slow-roll inflation it would remain curvature dominated and never form structure;  approximately 60 e-folds of slow-roll inflation are required to generate perturbations with an amplitude of approximately $10^{-5}$ and inflate away the curvature \cite{Freivogel:2005vv}.

\subsection{The model}

Computing CMB temperature fluctuations or density perturbations resulting  from the phase of eternal inflation is a difficult task because the perturbations are ${\cal O}(1)$ and cannot be treated perturbatively.  However, if eternal inflation is followed by a phase of ordinary slow-roll inflation with $\delta \rho/\rho \ll 1$ (\figref{pots}), the universe on small length scales can be described by a flat slicing plus perturbations.  If our patch of the universe has eternal inflation in its past, its large-scale homogeneity indicates that the eternal phase must have been followed by such an ordinary slow-roll phase, pushing the  ${\cal O}(1)$ fluctuations well outside the horizon.

Because of the many e-folds of the eternal inflation phase in the past, any memory of the initial conditions should have been erased, and therefore the fluctuations should be statistically homogeneous and isotropic, and Gaussian up to non-linear corrections.  Thus, they can be described by a power spectrum of fluctuations, $\mathcal{P}(| {\bf k}|=k)$, on a flat slice.  The metric should take the form
\be \label{met}
ds^2 = -(1-2 \Phi_N)dt^2 + a^2(t)(1+2\Phi_N)\delta_{ij}dx^idx^j,
\ee
where we have ignored anisotropic stress and dropped vector and tensor perturbations as subdominant.  Then for isotropic, homogeneous, and Gaussian fluctuations
\be \label{phinewt}
\Phi_N(\bold{x}) = \frac{1}{(2\pi)^\frac{3}{2}}\int d^3k ~g(\bold{k})e^{i\bold{k}\cdot\bold{x}}
\ee
where $g(\bold{k})$ is a Gaussian random variable with variance
\be \label{g}
\langle g^*(\bold{k})g(\bold{k'})\rangle_g = \frac{2\pi^2}{k^3}\mathcal{P}(k)\delta^3(\bold{k}-\bold{k'}),
\ee
and $\langle \cdot \rangle_g$ denotes an average over the distribution $g(\bold{k})$.

The power spectrum $\mathcal{P}(k)$ should become ${\cal O}(1)$ at the length scale $L_* = 2 \pi/k_*$ corresponding to the last mode that exited the horizon during the eternally inflating phase.  On shorter length scales it will be some smooth function that must approach $(10^{-5})^2$ on observable scales.  In other words,
	\be
	\mathcal{P}(k) = \left\{ \begin{array}{ll}
 \sim 1 & \textrm{for  $k \ll k_*$}\\ \sim 10^{-10} & \textrm{for $k  \gg k_*.$}\\ \end{array} \right.,
	\ee
Later in the paper (\secref{probsec}), we will take this transition to be sharp.  In general one could consider any function consistent with the above limits, but we expect that our results are very insensitive  to changes in the shape of the power spectrum.

As mentioned, FVEI requires a slightly separate treatment.  In FVEI, the amplitude of the perturbations generated in the false vacuum are set by its energy scale relative to the Planck mass, and may or may not be large.  Moreover, the exit from the false vacuum is believed to occur via the formation of a bubble with $SO(3,1)$ symmetry \cite{cdl}.  Inside this bubble is a homogeneous and isotropic universe with \emph{negative} curvature (see \emph{e.g.} \cite{Freivogel:2005vv}).  Applying the ansatz above to such a universe is possible, but it means that there is an isotropic contribution to $\Phi_N$ proportional to $(r/R)^2$, where $R$ is the radius of curvature.  After the tunneling, a period of slow-roll inflation is necessary to solve the flatness problem and seed structure \cite{Freivogel:2005vv}.  Therefore, there will be a Gaussian random component to $\Phi_N$ as well.  At the linear level, the isotropic component contributes in a simple way to the measured curvature, making it more negative.  Hence, an observation of \emph{positive} curvature would rule out this model at a confidence at least as high as it would for SREI.

\paragraph{A note on Hawking-Moss:}  In models where a broad barrier separates the false vacuum from ours, FVEI can come to an end via a transition first described by Hawking and Moss \cite{Hawking:1981fz}.  Originally, the Hawking-Moss instanton  was interpreted as a transition in which the entire universe tunnels to the top of the barrier as a de Sitter space, after which it rolls down, undergoes a phase of (ordinary) slow-roll inflation, and reheats.  Such a universe could have a \emph{positive} spatial curvature, despite originating from FVEI. \cite{Hawking:1981fz}  However, a more plausible interpretation of this instanton is instead that it computes the amplitude for a single Hubble volume to fluctuate to the top of the barrier.  Regardless, it turns out that the Hawking-Moss instanton only contributes to transitions when the top of the barrier is sufficiently broad to itself support SREI (in other cases, the Coleman-de Luccia instanton has a more negative action).  Therefore, transitions from the false vacuum mediated by Hawking-Moss are always followed by a phase of SREI \cite{1987PhLB..199..351A, batra}, to which our analysis applies.

\section{Curvature and the cosmic microwave background}

\subsection{Curvature}

Our definition of ``measured spatial curvature" will be the average of the Ricci scalar of \eqref{met} over a large spherical volume, which we will take to be the last scattering sphere.  While the various observational measures of $\Omega_k$ do not precisely correspond to this quantity, one of the most direct---the angular size of the first acoustic peak in the CMB power spectrum---is very closely related, and the others are expected to differ only to a degree determined by sub-horizon perturbations.  In other words, different observational measures of the spatial curvature will generically differ from each other and from our definition at the $10^{-5}$ level even given optimal statistics---but this accuracy more than suffices for our purposes.

Keeping only terms linear in $\Phi_N$,  the spatial Ricci curvature scalar $\mathcal{R}^{(3)}$ for the metric \eqref{met} is:
	\be \label{ricci}
	 \mathcal{R}^{(3)} = \frac{4 \nabla^2\Phi_N}{a^2}.
	 \ee
Using \eqref{phinewt}, calculating $\nabla^2\Phi_N$ is especially straightforward:
	\be
	\mathcal{R}^{(3)} =\frac{4}{a^2} \nabla^2\Phi_N = \frac{4}{a^2(2\pi)^\frac{3}{2}}\int d^3k ~g(\bold{k})(-k^2)e^{i\bold{k}\cdot\bold{x}}.
	\ee
Since $\langle g(\bold{k}) \rangle_g=0$, $\langle\mathcal{R}^{(3)}\rangle_g=0$. Taking the variance gives a non-zero result:
	\bea \label{riccivar}
	\langle|\mathcal{R}^{(3)}|^2\rangle_g &=& \frac{16}{a^4(2\pi)^3} \int d^3k d^3k' \langle g^*(k)g(k')\rangle_g k^2k'^2e^{i(\bold{k'}-\bold{k})\cdot\bold{x}} \nonumber\\
	&=& \frac{16}{a^4(2\pi)^3}\int d^3k \frac{2\pi^2}{k^3}\mathcal{P}(k)k^4 \nonumber \\
	&=& \frac{16}{a^4}\int dk\mathcal{P}(k) k^3
	 \eea
where in the second equality we have inserted \eqref{g} and used the delta function to eliminate $\bold{k'}$.

To compute the curvature averaged over a sphere of radius $R$ it is convenient to use a spherical decomposition of $\Phi_N$: 
	\be \label{phisph}
	\Phi_N(r,\theta,\phi)=\sqrt{\frac{2}{\pi}}\int dk \sum_{l,m}g_{lm}(k)kj_l(kr)Y_l^m(\theta,\phi)	,
	\ee
where the $j_l$ are the spherical Bessel functions and the $Y_l^m$ are the spherical harmonics.  Given \eqref{phinewt} and  \eqref{g} the $g_{lm}$ are Gaussian with mean zero and no covariance:
	\be \label{specsph}
	\langle g_{lm}^*(k)g_{l'm'}(k') \rangle = \frac{2\pi^2}{k^3}\mathcal{P}(k)\delta(k-k')\delta_{ll'}\delta_{mm'}.
	\ee
Using these coordinates we compute:
	\bea
	\mathcal{R}^{(3)}_{av} &=& \frac{\int_V\mathcal{R}^{(3)} dV}{V} \nonumber \\
	&=&\frac{4 \int_V \nabla^2\Phi_N dV}{\frac{4}{3}\pi R^3 a^{2}} \nonumber \\
	&=& \frac{3}{\pi a^2R^3}\int_{\partial V}\nabla\Phi_N\cdot\bold{n} ~dS.
	\eea
where the last line follows from Guass's Law, and $\bold{n}$ is the outward unit normal to the boundary surface $S = \partial V$.  Using (\ref{phisph}) we find:
	\bea
	\mathcal{R}^{(3)}_{av} &=& \frac{3}{\pi a^2R^3}\int d\theta d\phi R^2\sin{\theta} \frac{\partial}{\partial r}\!\left[\sqrt{\frac{2}{\pi}}\int dk \sum_{l,m}g_{lm}(k)kj_l(kr)Y_l^m(\theta,\phi)\right]_{r=R} \nonumber \\
	&=& \frac{6\sqrt{2}}{\pi a^2R}\int dk g_{00}(k)\frac{k}{R}\left(\cos{kR}-\frac{\sin{kR}}{kR}\right).
	\eea

Since $\langle g_{00}(k) \rangle_g=0$ it follows that $\langle\mathcal{R}^{(3)}_{av}\rangle_g=0$.  However, the variance of the spatially averaged curvature is non-zero;
	\bea \label{avericcivar}
	\langle \left( \mathcal{R}^{(3)}_{av} \right)^2\rangle_g&=&\frac{72}{\pi^2 a^4R^4}\int \!dkdk'\langle g_{00}^*(k)g_{00}(k')\rangle kk'\!\left(\cos{kR}-\frac{\sin{kR}}{kR}\right)\!\!\left(\cos{k'R}-\frac{\sin{k'R}}{k'R}\right)\nonumber \\
		&=& \frac{144}{a^4R^4}\int dk \frac{\mathcal{P}(k)}{k}\left(\cos{kR}-\frac{\sin{kR}}{kR}\right)^2.
	\eea
To translate these results into more familiar language, we can relate the Ricci 3-curvature of a standard curved FRW cosmology to $\Omega_k \equiv -{k}/{a^2H^2}$,
 where $k=1,0,-1$ corresponds to a spherical, flat or hyperbolic universe as usual.  Given the curved FRW metric $ds_3^2=a^2(t)\left( dr^2/(1-kr^2) + r^2 d\Omega_2^2 \right)$, the Ricci scalar is:
	 \begin{equation}
	 \mathcal{R}^{(3)}=\frac{6k}{a^2}.
	 \end{equation}
Thus $\Omega_k =-{\mathcal{R}^{(3)}}/{6H^2}$ and
	\be \label{omegaavvar}
	\langle \Omega_{k\, av}^2\rangle_g =\frac{ \langle \left( \mathcal{R}^{(3)}_{av} \right)^2\rangle_g}{36H^4}.
	\ee
If we evaluate (\ref{avericcivar}) at the time of last scattering $t_{LS}$ we will have:
         \be 
	\langle \Omega_{k\, av}(t_{LS})^2\rangle_g=\frac{4\int dk k^{-1}{\mathcal{P}(k)}\left(\cos{kR}-\frac{\sin{kR}}{kR}\right)^2}{a_{LS}^4H_{LS}^4R^4}.
	 \ee
Since $\langle\Omega_{k\, av}^2\rangle_g$ scales as ${(aH)}^{-4}$ we can relate $\langle|\Omega_{k\, av}(t_{LS})|^2\rangle_g$ to the value today via:
         \be \label{omegaav}
	 \langle \Omega_{k\, av}(t_0)^2\rangle_g=\frac{a^4_{LS}H^4_{LS}}{a^4_0H^4_0}\langle \Omega_{k\, av}(t_{LS})^2\rangle_g = {4 \over H_0^4R^4} {\int dk \frac{\mathcal{P}(k)}{k}\left(\cos{kR}-\frac{\sin{kR}}{kR}\right)^2},
	 \ee
where we have set $a_{0}=1$.  If $\mathcal{P}(k)$ has support mainly at $k \ll 1/R$, this becomes
 \be \label{omegask}
	 \langle \Omega_{k\, av}^2\rangle_g \approx {4 \over 9 H_0^4} {\int_{0}^{1/R} dk ~ k^{3}{\mathcal{P}(k)}}.
	 \ee
\subsection{CMB Multipoles}

In the Sachs-Wolfe approximation \cite{sw, huwhite}, the CMB temperature perturbations are related to the Newtonian potential on the last scattering sphere by
	\be
	\frac{\delta T}{T_{0}}(\theta, \phi) = -\frac{1}{3}\Phi_N (R_{LS}, \theta, \phi, t_{LS}),
	\ee
where $T_{0}$ is the background temperature of the CMB (2.725 K), and $R_{LS}$ is the comoving distance to the edge of the earth's past lightcone at the last scattering time $t_{LS}$.  

The temperature anisotropy map, $\delta T(\theta,\phi)$, is typically decomposed using spherical harmonics:
	\be
	a_{lm} \equiv \int d\theta d\phi \sin{\theta} {Y_l^m}^* \delta T .
	\ee
The $l^{th}$ multipole moment of the temperature anisotropy is 
\be	
	C_l\equiv \frac{1}{2l+1}\sum_{m=-l}^l a_{lm}^*a_{lm}.
	\ee
Because of isotropy, $\langle a_{lm}^*a_{lm}\rangle_g$ is independent of $m$:
	\be
	C_l= \langle a_{lm}^*a_{lm}\rangle_g.
	\ee
	
	In terms of $\Phi_N$ we have:
	\bea
	a_{lm} &=&-\frac{T_{0}}{3}\int d\theta d\phi \sin{\theta} \Phi_N(R_{LS},\theta,\phi) Y_l^{m*}(\theta,\phi) \nonumber \\
	&=&-\frac{T_{0}}{3}\sqrt{\frac{2}{\pi}}\int d\theta d\phi \sin{\theta} dk \sum_{l',m'} g_{l'm'}(k)kj_{l'}(kR_{LS})Y_{l'}^{m'}(\theta,\phi)Y_l^{m*}(\theta,\phi) \nonumber \\
	&=& -\frac{T_{0}}{3}\sqrt{\frac{2}{\pi}}\int dk g_{lm}(k)kj_l(kR_{LS}).
	\eea
	Therefore the $l^{th}$ multipole is:
	\bea \label{multi}
	\langle C_l \rangle_g &=& \frac{2T_{0}^2}{9\pi}\int dk dk' \langle g_{lm}^*(k)g_{l'm'}(k')\rangle_gkk'j_l(kR_{LS})j_l(k'R_{LS}) \nonumber \\
	&=& \frac{2T_{0}^2}{9\pi}\int dk \frac{2\pi^2}{k^3}\mathcal{P}(k)k^2j_l^2(kR_{LS}) \nonumber \\
	&=& \frac{4\pi T_{0}^2}{9}\int dk \frac{1}{k}\mathcal{P}(k)j_l^2(kR_{LS}).
	\eea
In the case of the quadrupole $l=2$, the $k \ll 1/R_{LS}$ part of the integral is
\be \label{quadapprox}
T_{0}^{-2} \langle C_{2} \rangle_g  \approx {4 \pi R_{LS}^{4} \over 2025} \int dk ~ k^{3}{\mathcal{P}(k)}
\ee
Comparing \eqref{quadapprox} to \eqref{omegask} shows that long wavelength modes contribute to the variance in the average spatial curvature at almost the same level they compute to the quadrupole: 
\be \label{compare}
{T_{0}^{-2} \langle C_2 \rangle \over  \langle \Omega_{k}^2\rangle } \approx 1.7,
\ee
where we have used $R_{LS}\approx 3.3 H_{0}^{-1}$ \cite{2006PASP..118.1711W}.

\section{Predicting curvature}

In this section we will first sketch the constraints that the observed CMB multipoles place on the spatial curvature by considering two extreme examples.  Then we will calculate a probability distribution for the spatial curvature from a conservative simplification of a power spectrum that would arise from either SREI or FVEI.

\subsection{Flat spectrum}\label{flat}

We first consider an exactly scale invariant power spectrum
	\be
	\mathcal{P}_0(k) = \epsilon.
	\ee
This is a fairly good approximation to the power spectrum inferred from observation, which has a tilt $1-n_s=.04 \ll 1 $.\cite{Komatsu:2010fb}  Matching to the WMAP results, 
$$\epsilon \approx 1.273\times 10^{-9}.$$

With this scale-invariant spectrum, the variance of the Ricci scalar (\ref{riccivar}) has a $k^4$ ultraviolet divergence (due to the fact that we are including curvature on arbitrarily small length scales).  The variance of the spatially averaged curvature (\ref{avericcivar}) is better behaved:
	\bea
	\langle|\mathcal{R}^{(3)}_{av}|^2\rangle_g &=& \frac{1.8\times10^{-7}}{a^4R^4}\int_0^K dk\frac{1}{k}\left(\cos{kR}-\frac{\sin{kR}}{kR}\right)^2 \nonumber \\
	&=& \frac{1.8 \times 10^{-7}}{a^4R^4} \left[ .14 + \ln{\sqrt{K R}} + {\cal O}(1/KR) \right],
	\eea
where $K$ is a cutoff on large momentum.  The cutoff $K \sim 2\pi H_{eq}$, the inverse horizon size at matter-radiation equality where the transfer function for $\Phi_N$ falls as $k^{-2}$ \cite{weinbook}.
Setting $K=2\pi H_{eq}\approx6.2\times10^5H_0$, averaging over a sphere of $R=R_{LS}\approx 3.3H_0^{-1}$ and using \eqref{omegaavvar} gives:
$$
\sqrt{\langle\Omega_{k\, av}^2\rangle} \approx 9.5\times10^{-6}.
$$

This estimate is rough and probably too low, but it demonstrates that  sub-horizon perturbations ${\mathcal P}(k)\sim 10^{-10}$ do not contribute to $\langle \Omega_k \rangle$ above the level of $\sim 10^{-5}$.  That leaves open the possibility that super-horizon perturbations might contribute to it.

\subsection{High power on large scales}\label{high power}
Next, we consider a universe in which there is no power up to some scale $L_*$, and then fluctuations are ${\cal O}(1)$ on scales greater than $L_*$.  In momentum space this corresponds to a primordial power spectrum
	\be \label{stepspec}
	\mathcal{P}=\beta ~\theta(k_*-k)
	\ee
where $\beta$ is the $\mathcal{O}(1)$ parameter, $k_*=\frac{2\pi}{L_*}$, and $\theta$ is a step function.  

Using (\ref{multi}) we compute the first 3 multipole moments:
	\bea \label{stepquad}
	C_2 &=& \frac{4\pi T_{0}^2\beta}{9}\int_0^{k_*}dk\frac{j_2^2(kR_{LS})}{k}\nonumber \\
	&= & \frac{\pi T_{0}^2\beta(k_*R_{LS})^4}{2025} -\frac{2\pi T_{0}^2\beta (k_*R_{LS})^6}{42525}+\frac{2\pi T_{0}^2\beta(k_*R_{LS})^8}{893025}+\mathcal{O}(k_*R_{LS})^{10},\\
	\nonumber\\
	C_3 &=& \frac{4\pi T_{0}^2\beta}{9}\int_0^{k_*}dk\frac{j_3^2(kR_{LS})}{k}\nonumber \\
	&=& \label{stepoct} \frac{2\pi T_{0}^2\beta(k_*R_{LS})^6}{297675} - \frac{\pi T_{0}^2\beta (k_*R_{LS})^8}{1786050} + \mathcal{O}(k_*R_{LS})^{10}\\
	\nonumber\\
	C_4 &=& \frac{4\pi T_{0}^2\beta}{9}\int_0^{k_*}dk\frac{j_4^2(kR_{LS})}{k}\nonumber \\
	&=& \frac{\pi T_{0}^2\beta (k_*R_{LS})^8}{16074450} + \mathcal{O}(k_*R_{LS})^{10}.
	\eea

Higher multipoles are suppressed by increasing powers of $R_{LS}/L_*$, and as such are irrelevant for constraining $L_*$.  (Intuitively, this is because a long wavelength $k$ mode doesn't overlap much with a high $l$ multipole moment.)  Only these first three multipoles receive appreciable corrections from long wavelength modes, and they are all we consider in the analysis of \secref{probsec}.

Using \eqref{stepspec} in \eqref{omegaav} we find:
	\bea \label{ro}
	\langle \Omega_{k\, av}^2\rangle_g &=& \frac{4 \beta}{H_0^4R^4}\int_0^{k_*}{ dk \over k}{\left(\cos{kR}-\frac{\sin{kR}}{kR}\right)^2} =
	\frac{\beta}{(H_0R)^4(k_*R)^2} \Big\{\cos{2k_*R}-1  \nonumber \\
	 & + & 2(k_*R)^2  (\gamma-1-{\rm Ci}(2k_*R)+\log{2k_*R}) +2k_*R\sin{2k_*R}\Big\} \nonumber \\
	 &=&  {\beta k_{*}^{4} \over 9 H_{0}^{4}} - {2 \beta R^2 k_{*}^{6} \over 135 H_{0}^{4}} + {\cal O}(k_{*}^{8}).
	\eea
Comparing to \eqref{stepquad} (recall that $R_{LS} \approx 3.3 H_0^{-1}$) shows that the variance in the averaged curvature depends on the long wavelength perturbations in almost the same way as the variance in the quadrupole, as already anticipated for general ${\mathcal P}(k)$ in \eqref{compare}.  Therefore, measuring $\Omega_{k}^2 \gg C_{2}/T_{0}^{2}$ is very unlikely in this model.

\subsection{Probability of $\Omega_k$ from eternal inflation}\label{probsec}

A rough model for the power spectrum of perturbations generated by an eternally inflating phase followed by a phase of ordinary slow roll inflation is the combination of the two cases considered in the preceding subsections:
	\be \label{ps}
	\mathcal{P}(k) = \left\{ \begin{array}{ll}
\beta & \textrm{for  $k < k_*$}\\ 1.3\times 10^{-9} & \textrm{for $k > k_*$}\\ \end{array} \right.,
	\ee
where $\beta$ is an $\mathcal{O}(1)$ parameter.  The power on small scales has been normalized so that in the limit of large $L_*$ the octopole matches the output of CAMB \cite{Lewis:1999bs} with the WMAP 7 year bet-fit parameters  \cite{Komatsu:2010fb}, except with the scalar spectral index set to one.

In the subsequent analysis we renormalize $C_2$ and $C_4$  by hand so that in the large $L_*$ they match the CAMB output.  This incorporates sub-dominant effects like the integrated Sachs-Wolfe  into our results (which in any case are insensitive to such details).

The sharp transition model \eqref{ps} is a conservative choice in the sense that additional power on scales between the last scattering length and  $L_*=2 \pi/k_*$, the scale where the power is ${\cal O}(1)$, would serve to push $L_*$ further out given the observational constraints.  That in turn makes the curvature even less likely to be large.  In any case, our results for the confidence are so robust that they are almost entirely insensitive to changes in the shape (or even amplitude, see below) of the power spectrum at large scales.  As \eqref{compare} demonstrates, the specific form of ${\mathcal P}(k)$ is not relevant.  All that matters is that the power is large at some large scale $L_{*}$.

The value for $\Omega_{k\, av}$ and the CMB multipoles will depend on the scale $L_*$, which we integrate out:
	\begin{equation}
	P\left( \Omega_{k\, av} | C_2, C_3,C_4\right) = \int dL_*  P\left(\Omega_{k\, av}, L_* | C_2, C_3,C_4\right).
	\end{equation}
Using $P\left(A|B\right) ={P\left(A,B\right)}/{P(B)}$ and Bayes's theorem 
	$P(A|B)={P(B|A)P(A)}/{P(B)}$:
	\begin{eqnarray}
	P\left( \Omega_{k\, av} | C_2, C_3,C_4\right) &=& \int  dL_* \, P\left(\Omega_{k\, av}, L_* | C_2, C_3,C_4\right)  \nonumber\\
	&=& \int dL_*\, P(\Omega_{k\, av}|L_*,C_2,C_3,C_4)P(L_*|C_2,C_3,C_4) \nonumber\\
	&=& \int dL_* \, P(\Omega_{k\, av}|L_*)\frac{P(C_2,C_3,C_4|L_*)P(L_*)}{P(C_2,C_3,C_4)} \nonumber\\
	&=& \int dL_* \, P(\Omega_{k\, av}|L_*)\frac{P(C_2|L_*)  P(C_3|L_*)P(C_4|L_*)P(L_*)}{P(C_2,C_3,C_4)}.
	\end{eqnarray}
We have used the fact that \emph{e.g.} $P(\Omega_{k\, av}|C_l,L_*) = P(\Omega_{k\, av}|L_*)$, which holds because the dependance of $\Omega_{k\, av}$ on $C_l$ is only through their mutual dependence on $L_*$.  The probability distribution $ P(\Omega_{k\, av}|L_*)$ is Gaussian, while $P(C_l|L_*)$ are $\chi^{2}$ distributed.

Using the same techniques we are also able to calculate the normalization factor $P(C_2,C_3,C_4)=\int dL_* P(C_2|L_*)	P(C_3|L_*)	P(C_4|L_*)	P(L_*)$.

	For the prior $P(L_*)$ we tried both a flat and a logarithmic distribution.  The results were very similar, and we present the (more conservative for our claim) log prior in what follows.
	
\section{Results}
The model defined by the power spectrum \eqref{ps} and a flat background \eqref{met} predicts a symmetric distribution for the measured spatial curvature $\Omega_{k\, av}$.  As discussed above this is an accurate description for universes that underwent SREI, but for universes that underwent FVEI, the measured value of the curvature will be shifted towards {negative} spatial curvature (more positive $\Omega_k$) by the globally negatively curved geometry of the bubble when it nucleates.  Because observation already constrains $|\Omega_k| < 10^{-2}$, this shift is linear and can be treated as an additive contribution to the local curvature $\Omega_{k\, av}$.  Therefore, FVEI models are falsified by measurements of \emph{positive} curvature ($\Omega_k < 0$) at an equal or greater level of confidence as the flat model \eqref{ps}.  

\begin{figure}[!htb]
\begin{center}
\includegraphics[width=\textwidth]{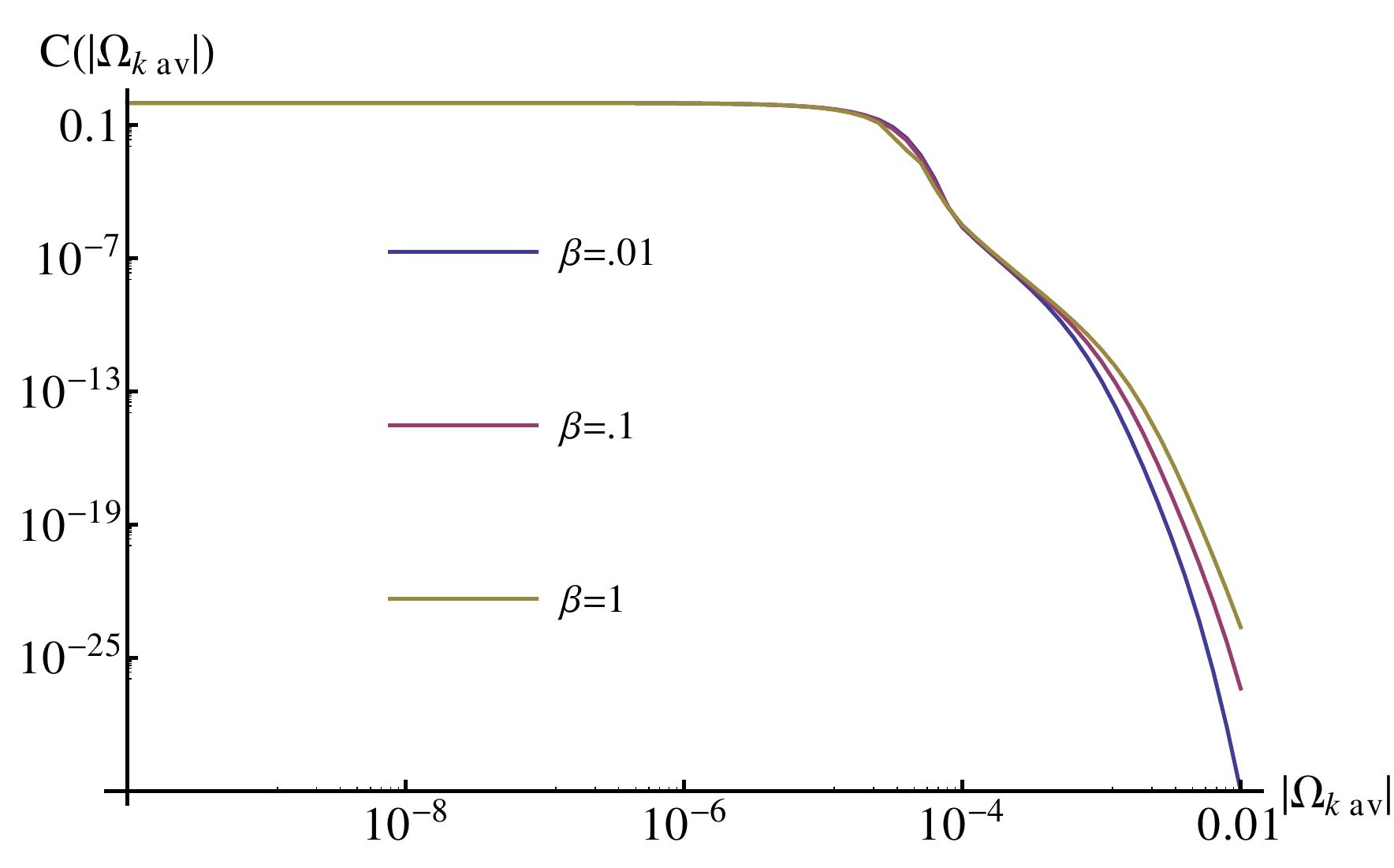}
\caption{\label{probdist}
 The cumulative probability distribution $C(|\Omega_k|)$ defined in \eqref{C} (the probability that the magnitude of the measured curvature is larger than $|\Omega_{k, av}|$), using the power spectrum \eqref{ps}, and a logarithmic prior on the eternal inflation length scale $L_*$.}
\end{center}
\end{figure}
The only parameter left undiscussed is $\beta$, the amplitude of the power spectrum on large scales  \eqref{ps}.  Because we have used linear perturbation theory, our results for  $\beta \simgeq 1$ are subject to ${\cal O}(1)$ corrections.  We tested a range of values $.01<\beta<2$, finding that our results are qualitatively very insensitive to $\beta$ in the regime of interest $10^{-4} < |\Omega_k| < 10^{-2}$.   Three representative values are plotted in \figref{probdist}.

Our main result is the cumulative probability distribution 
\be \label{C}
C(|\Omega_k|) \equiv 1- \int_{-|\Omega_k|}^{+|\Omega_k|} d \Omega_k \, P(\Omega_k)
\ee
(\figref{probdist}), the probability in the model \eqref{ps} that the magnitude of the measured curvature will be greater than $|\Omega_k|$.  

The probability drops very sharply for $|\Omega_{k\, av}| > 10^{-4}$, which (not coincidentally) coincides with the threshold reported in \cite{Vardanyan:2009ft} for a 5-$\sigma$ discovery of spatial curvature.  This establishes the claim made in the abstract and introduction.

While our analysis was done with a specific model \eqref{ps} for the power on large scales, this is a conservative choice in the sense that it \emph{reduces} the confidence with which one could falsify eternal inflation.  In any case, \eqref{compare} demonstrates that our conclusions are quite insensitive to the detailed form of ${\mathcal P}(k)$.

\section{Conclusions}

Scientific theories must be falsifiable, and the more potential falsifications they pass, the more confidence we should have in them.  In the near future, measurements of the spatial curvature are expected to improve by an order of magnitude, and future measurements of the 21cm transition could improve the constraints to the $10^{-4}$ level \cite{Mao:2008ug, Barenboim:2009ug}.  Thus, the near and medium term may see two full orders of magnitude improvement  in the constraints on curvature, making this test quite sharp.  

It is worth pointing out that our analysis is limited in several potentially important respects:
\begin{itemize}
\item We have treated the perturbations linearly, which is not adequate when $\beta \simgeq 1$.  However, it is difficult to see how non-linearities could affect the curvature---which is essentially the isotropic component of the quadrupole---and not the measured values of $a_{2m}$.
\item Strictly speaking, a measurement of negative spatial curvature would not falsify \emph{any} period of SREI in our past, if there was an intervening period of FVEI.\footnote{We thank B. Freivogel for this nitpick.}
\item The exit from FVEI by bubble nucleations leads to negative curvature.  In principle, another exit mechanism could exist that introduces positive curvature\footnote{As we have already noted, the Hawking-Moss transition does \emph{not} suffice for this.}    or differs from the model studied here in some other relevant respect.\footnote{An instructive example is a transition from a phase of lower dimensional FVEI, where the radion field(s) corresponding to one (two) compact dimensions tunnels from a minimum and begins to expand.\cite{Graham:2010hh, BlancoPillado:2010uw, Adamek:2010sg}  Because the resulting universe is anisotropic on large scales, the CMB naturally has a large quadrupole component.  The low measured value of $C_2$ is the best constraint on such models, for precisely the reasons discussed in this paper.}
\item A full understanding of eternal inflation remains a distant goal, and could require a new interpretation of quantum mechanics.\cite{Bousso:2011up}
\end{itemize}

As mentioned in the introduction, modulo a choice of measure an observation of curvature could falsify not just models that allow eternal inflation and in which the classical initial conditions  favor it, but all models that can eternally inflate in \emph{any} regime.  Most inflationary models are effective field theories, and most of them support eternal inflation somewhere in their field space.  More fundamentally, theories such as string theory appear to support eternal inflation in many different metastable minima.  As such, a measurement of positive spatial curvature could be very challenging to accommodate within the current theoretical framework in high-energy physics.

\section*{Acknowledgements}
We thank Guido D'Amico, Ben Freivogel, Alan Guth, Lam Hui, Chao-Lin Kuo, Roman Scoccimarro, and I-Sheng Yang for useful discussions.
MK is supported by NSF CAREER grant PHY-0645435.

\bibliographystyle{klebphys2}

\bibliography{bubbles}

\end{document}